\documentclass[pdflatex,sn-basic]{sn-jnl}

\usepackage{enumitem}            

\usepackage{amsmath}             
\usepackage{booktabs}            
\usepackage{tabularx}            
\usepackage{longtable}           
\usepackage{array}              
\usepackage{adjustbox}           
\usepackage{pdflscape}           

\usepackage{graphicx}            
\usepackage{subcaption}          
\usepackage{caption}             
\usepackage{listings}            
\usepackage{xcolor}              
\usepackage{colortbl}
\usepackage{tcolorbox}
\tcbset{
  colback=white,
  colframe=blue!40!black,
  coltitle=blue!20!black,
  fonttitle=\bfseries,
  boxsep=2pt,
  arc=3pt
}

\definecolor{headerblue}{RGB}{220,230,255}
\definecolor{rowgray}{RGB}{245,245,245}

\usepackage{tikz}                
\usetikzlibrary{positioning, arrows.meta} 

\theoremstyle{thmstyleone}

\theoremstyle{thmstyletwo}

\theoremstyle{thmstylethree}

\begin{document}

\title[Article Title]{Learning-Infused Formal Reasoning: From Contract Synthesis to Artifact Reuse and Formal Semantics}

\author*[1]{\fnm{Arshad} \sur{Beg}}\email{arshad.beg@mu.ie}
\equalcont{These authors contributed equally to this work.}

\author[1]{\fnm{Diarmuid} \sur{O'Donoghue}}\email{diarmuid.odonoghue@mu.ie}
\equalcont{These authors contributed equally to this work.}

\author[1]{\fnm{Rosemary} \sur{Monahan}}\email{rosemary.monahan@mu.ie}
\equalcont{These authors contributed equally to this work.}

\affil*[1]{\orgdiv{Department of Computer Science}, \orgname{{Maynooth University}, \country{Ireland}}}

\abstract{
This paper articulates a long-term research vision for formal methods at the intersection with artificial intelligence, outlining multiple conceptual and technical dimensions and reporting on our ongoing work toward realising this vision. It advances a forward-looking perspective on the next generation of formal methods based on the integration of automated contract synthesis, semantic artifact reuse, and refinement-based theory. We argue that future verification systems must builds towards individual correctness proofs toward a cumulative, knowledge-driven paradigm in which specifications, contracts, and proofs are continuously synthesised and transferred across systems. To support this shift, we outline a hybrid framework combining large language models with graph-based representations to enable scalable semantic matching and principled reuse of verification artifacts. Learning-based components provide semantic guidance across heterogeneous notations and abstraction levels, while symbolic matching ensures formal soundness. Grounded in compositional reasoning, this vision points toward verification ecosystems that evolve systematically, leveraging past verification efforts to accelerate future assurance.
}
\keywords{Formal Reasoning, Formal Verification, Contract Synthesis, Graph Construction, Graph Matching, Unifying Theory of Programming, Theory of Institutions}
\maketitle
\section{Introduction}

Over the past decade, AI has advanced at an incredible pace. Deep learning and large language models now handle language, images, and decision-making with impressive fluency. Yet despite this progress, these systems remain hard to trust. They are opaque, they can fail without warning, and they offer no clear guarantees of safety or correctness--problems that matter deeply in high-stakes settings. Formal methods offer something AI currently lacks: solid, mathematical guarantees about how a system behaves. They support engineers in writing precise rules and providing proofs that a system follows them. However, these techniques struggle with today’s large and constantly changing AI systems:specifications are difficult to maintain, proofs grow quickly in complexity, and symbolic tools alone cannot scale to modern learned components.

Recent research hints at a way forward. Learning can speed up proof search, propose useful specifications, and guide testing. Formal reasoning, in turn, is being extended to handle neural networks and other learned models.  However, many current solutions target narrow problems, missing is a unified foundation. Our vision, which we call Learning-Infused Formal Reasoning (LIFR), aims to bring these solutions together, by  creating a steady feedback loop where learning helps automate and adapt formal reasoning tasks, while formal methods ensure that the results remain sound and reliable. Techniques like automated contract synthesis and LLM-guided discovery of verification artifacts become part of a common framework supported by rigorous semantics. By integrating learning, reuse, and formal foundations, we will move beyond \textit{ad hoc} tool combinations toward a principled way of building intelligent systems that are both powerful and trustworthy. 

Now we present our recent contributions. The paper \cite{BegEtAl2025} offers a structured look at the challenges of turning informal, natural-language requirements into precise formal specifications. It highlights how ambiguity, missing domain models, contextual gaps, and the instability of current LLMs jointly make consistent and reliable formalisation difficult. A key contribution is the introduction of the VERIFYAI research framework, which proposed to integrate LLMs with NLP pipelines, ontology-driven modelling, and artefact-reuse mechanisms to support semi-automated derivation of verifiable specifications. The paper also offers a focused synthesis of early state-of-the-art approaches--such as Req2Spec, SpecGen, AssertLLM, and nl2spec--systematically comparing how contemporary systems represent requirements, generate logical constraints, and interface with verification tools. Additionally, it reports empirical evaluations across multiple SMT solvers (Alt-Ergo, Z3, CVC4, CVC5) using Frama-C PathCrawler, providing comparative evidence on verification success rates and solver execution times that highlight practical feasibility limits of current formalisation pipelines.

The more comprehensive study \cite{Beg2025Traceable} significantly extends these contributions by offering a large-scale synthesis of over one hundred studies that examine AI-enabled approaches to requirements formalisation, traceability, and verification.  This work  maps the broader landscape of LLM-based and AI-assisted techniques for translating informal requirements into formal specifications across diverse programming languages, specification notations, and verification backends, including model checkers and theorem provers. The survey categorises the dominant methodological patterns used in recent tools. These patterns range from pattern-based rule extraction to logical constraint inference and hybrid neuro-symbolic pipelines. The survey also identifies ecosystem-level challenges that remain largely unresolved. These challenges include barriers to toolchain integration, insufficient traceability mechanisms, and limited reuse of existing specification artefacts.

The work incorporates an expanded empirical investigation of Frama-C's formal analysis and test-generation ecosystem. The study evaluates the Runtime Error (RTE) analyser, the Value Analysis (EVA) plugin, and the PathCrawler tool to determine their capability to infer assertions, generate ACSL annotations, detect runtime anomalies, and produce targeted test cases for representative C programs. The findings reveal persistent issues--including solver instability, path-coverage constraints, and configuration sensitivities--that constrain automation reliability in real-world formalisation workflows. By proposing mitigation strategies such as dataset adaptation, selective path-exploration heuristics, and fine-grained tool-configuration adjustments, it proposes how coordinated, multi-tool, human-guided workflows can systematically overcome these limitations. Together, these two contributions \cite{BegEtAl2025, Beg2025Traceable} provide foundational exploration and experimental validation of AI-enabled formal methods.

\begin{figure}[h]
    \centering
    \includegraphics[width=0.6\linewidth, keepaspectratio]{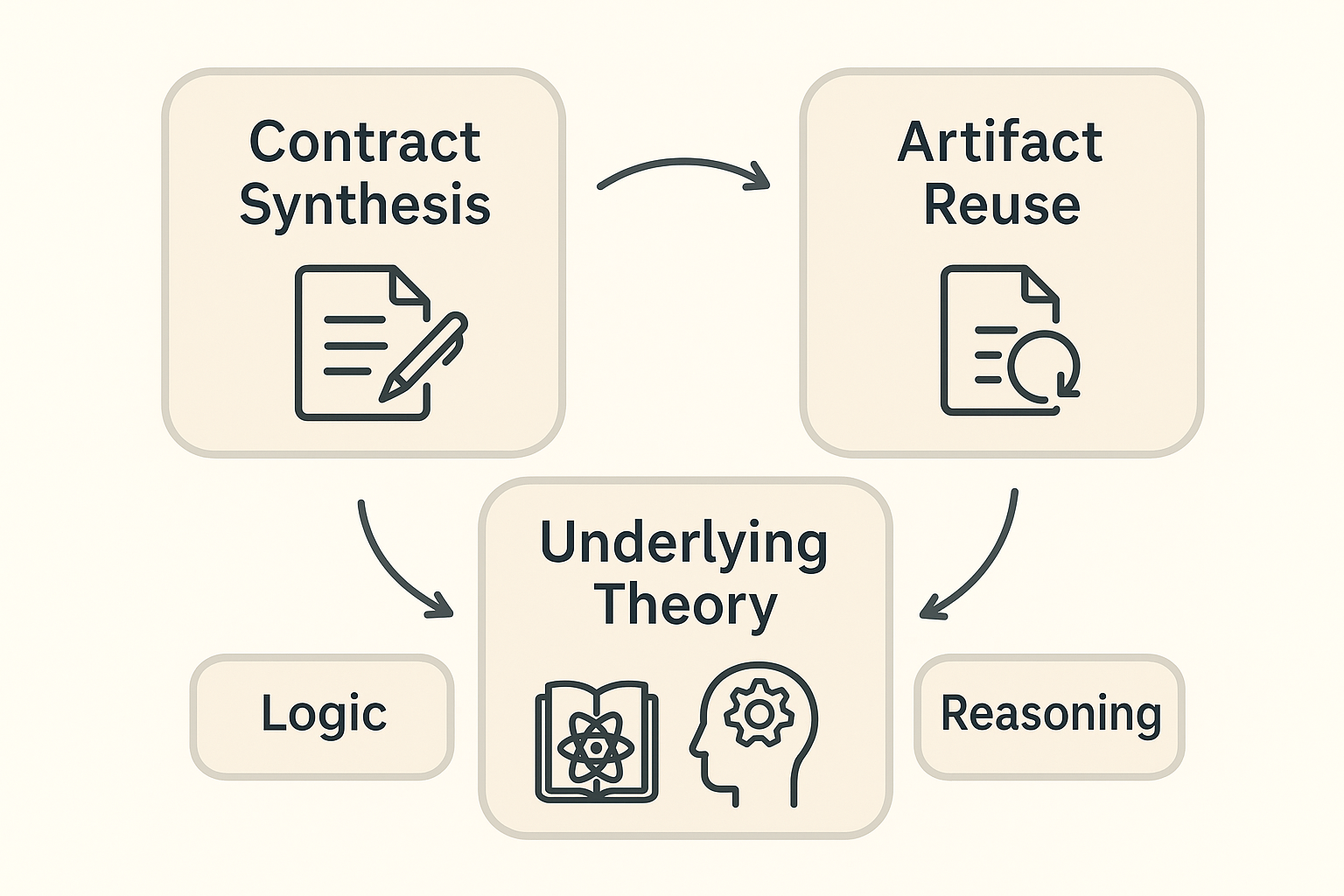}
    \caption{High-level conceptual architecture of the LIFR framework.}
    \label{fig:conceptualframework}
\end{figure}

Our research vision comprises of three threads: contract synthesis, artefact reuse, and semantic foundations. Figure~\ref{fig:conceptualframework} presents the high-level conceptual architecture that organises the research threads into the unified LIFR framework.

The structure of the paper is as follows: Section \ref{sec:contractsynthesis} presents our vision of research in the area of contract synthesis. Subsection \ref{subsec:empirical-contract-synthesis} briefs our contribution addressing contract synthesis. Section \ref{sec:graphmatching} explores the role of LLMs in graph matching for artefact reuse. Subsection \ref{subsec:graph-construction-matching} summaries our on-going work, while Section \ref{sec:underlyingtheory} describes the formal semantic foundations. Section \ref{sec:discussion} summarises the challenges and future directions identified in \cite{Beg2025Traceable}. Section \ref{sec:conclusions} concludes the paper.

\section{Contract Synthesis}
\label{sec:contractsynthesis}

This section articulates a research vision for contract synthesis situated within the proposed VERIFYAI framework \cite{BegEtAl2025}. Rather than presenting a completed implementation, we outline a forward-looking architectural blueprint for how large language models (LLMs), verification engines, and human-centred oversight could be combined to support semantically rigorous specification workflows. The aim is to characterise the methodological foundations required for transforming natural-language requirements into verifiable program contracts while avoiding assumptions of system maturity or deployment. In contrast to the artifact reuse setting of Section~\ref{sec:graphmatching}, which focuses on semantic alignment between existing formal artefacts, the present section concentrates on the initial construction of contracts from informal descriptions and on the principles needed to ensure their correctness, traceability, and interpretability. The following subsections develop this vision by analysing the challenges of requirement formalisation, motivating neuro-symbolic synthesis, and describing the intended architectural components of VERIFYAI as a conceptual pipeline.

\subsection{From Natural Language to Verifiable Contracts}

The proposed VERIFYAI framework is motivated by long-standing difficulties in converting informal requirements into precise, machine-verifiable annotations. Prior studies \cite{BegEtAl2025, Beg2025Traceable} highlight that natural-language descriptions often contain latent assumptions, contextual dependencies, and domain-specific conventions that are not directly compatible with the rigid semantic formats required by formal verification environments. LLMs provide new opportunities for interpreting such descriptions, but their outputs are susceptible to ambiguity, instability, and hallucination, especially when required to produce logically constrained structures such as preconditions, postconditions, and invariants.

Within VERIFYAI, contract synthesis is conceived as an iterative transformation process rather than a direct translation. The framework envisions the use of controlled prompting strategies that scaffold LLMs to extract candidate behavioural obligations from textual descriptions. These candidates would then be subjected to validation and refinement steps informed by formal verification feedback. Rather than assuming that LLMs can inherently internalise program semantics, this process emphasises the need for explicit alignment between linguistic intent and symbolic constraints. The intended outcome is a synthesis workflow in which LLM-generated artefacts converge toward logical adequacy through repeated interaction with verification tools, while maintaining fidelity to the original requirements.

\subsection{Architectural Foundations of Learning-Infused Contract Synthesis}

The architectural sketch underpinning VERIFYAI comprises three complementary technical pillars, each grounded in insights from the literature but intentionally presented as a conceptual proposal rather than a realised system.

The first pillar concerns a structured prompting architecture shaped by the empirical observations summarised in \cite{BegEtAl2025}. These observations suggest that LLM performance in specification-related tasks is heavily influenced by prompt design, contextual cues, and the use of domain templates. VERIFYAI extends these findings by outlining a family of prompt patterns and transformation routines intended to normalise natural-language requirements, expose implicit semantic relations, and guide the generation of candidate program contracts. This layer establishes the linguistic and syntactic scaffolding needed for subsequent verification-aware refinement.

The second pillar integrates formal verification into the generative loop. Verification tools such as Frama-C, Z3, or Alt-Ergo are envisioned as active components that provide structured feedback on the correctness and completeness of the generated annotations. Counterexamples, proof failures, and satisfiability results would be used to reshape and constrain subsequent LLM outputs, thereby counteracting hallucinations and enforcing semantic discipline. This neuro-symbolic interplay conceptualises contract synthesis as a convergence process driven jointly by generative heuristics and symbolic reasoning, without presupposing that such integration is currently automated.

The third pillar focuses on traceability and expert oversight, drawing on the limitations identified in \cite{Beg2025Traceable}. Current AI-assisted formalisation approaches frequently lack mechanisms for documenting how generated contracts relate to initial requirements, which complicates auditability and expert review. VERIFYAI proposes a design in which mappings between requirements, generated annotations, and revision histories are explicitly recorded. This would support human-in-the-loop curation, facilitate accountability, and enable engineers to understand and shape the evolution of the specification artefacts. The purpose of this pillar is not to prescribe a finished interface, but to articulate the metadata and interaction patterns necessary for trustworthy adoption in safety-critical domains.

\subsection{Scalable and Semantically Grounded Contract Synthesis}

The VERIFYAI proposal anticipates that contract synthesis must ultimately operate across diverse languages, verification platforms, and specification paradigms. This end-points of specification distinguish contract construction from the artifact reuse mechanisms discussed in Section~\ref{sec:graphmatching}. Here, the goal is not to discover correspondences between pre-existing artefacts, but to ensure that newly synthesised contracts can integrate flexibly into heterogeneous verification ecosystems.

To this end, the conceptual architecture is intentionally language and tool-agnostic. It is designed to accommodate ACSL-style annotations, SMT-based constraints, or specifications for model checkers and theorem provers. This generality is informed by limitations noted in \cite{Beg2025Traceable}, where vertically siloed toolchains hinder interoperability and long-term maintainability. By foregrounding semantic grounding and verifier-driven refinement, VERIFYAI aims to establish an extensible basis from which specialised contract styles can be generated without compromising correctness.

The long-term research direction implied by this architecture is that contract synthesis may act as a foundational capability for a broader family of learning-enhanced formal reasoning tasks. Synthesised contracts could serve downstream roles in model checking, symbolic execution, testing, or behavioural refinement, provided that their semantics are precisely defined and traceable. Although VERIFYAI is presented purely as a proposal, its envisioned integration of synthesis, verification, and expert oversight outlines what such scalable and semantically principled workflows may eventually require.

\subsection{Empirical Advances in Contract Synthesis}
\label{subsec:empirical-contract-synthesis}

 \cite{ftfjp2026acslWithLLMs} investigates the practical behaviour of automated contract synthesis for C programs through the generation of ACSL annotations. The study evaluates five approaches: our own rule-based Python script, the Frama-C RTE plugin, and three large language models (DeepSeek, GPT-5, and OLMo3). The focus is on one-shot annotation generation, where specifications are produced without iterative refinement. All generated annotations are verified using the Frama-C WP plugin with multiple SMT solvers under a controlled setup. A filtered subset of the CASP benchmark is used to ensure consistency across experiments. The evaluation considers proof success rates, solver timeouts, and internal processing time. This design allows a direct comparison between symbolic and learning-based approaches while isolating the effect of annotation quality. The results provide a clear empirical basis for understanding the current capabilities and limitations of automated contract generation methods.

The findings show that rule-based approaches and the RTE plugin provide consistent verification performance, particularly for safety-related properties such as runtime checks. At the same time, LLM-based approaches demonstrate competitive effectiveness, with verification success rates remaining in the range of 83--95\% across models. In addition to this strong baseline performance, LLMs generate a broader range of annotations, including functional constraints that are not typically captured by rule-based techniques. While some variability is observed in solver behaviour and processing time, this reflects differences in the structure and richness of generated specifications rather than a lack of capability. Differences across models are also evident, with DeepSeek exhibiting relatively stable behaviour, though all evaluated models produce practically useful annotations. Overall, the results indicate that LLM-based methods are viable for contract generation and can be meaningfully integrated into verification workflows.

These observations have direct implications for the contract synthesis proposed in this paper. They support the view that synthesis should be treated as a guided process where multiple generation strategies contribute complementary strengths. Rule-based methods provide reliable coverage for safety properties, while LLMs extend this coverage by introducing richer and more expressive candidate constraints. This combination enables broader specification discovery without sacrificing verification feasibility. Within the LIFR framework, this aligns with a neuro-symbolic loop in which LLM-generated contracts are validated and refined through formal verification. Practitioners can therefore adopt LLM-based annotation generation as part of their workflow, particularly in contexts where specification effort is a bottleneck. The results strengthen the case for integrating learning-based generation with symbolic reasoning to build scalable and effective contract synthesis systems.

\section{LLMs and Graph Matching for Artifact Reuse}
\label{sec:graphmatching}

This section lays out a research vision for reuse of verification artifacts with the help of large language models (LLMs) and graph matching. It focuses on how these techniques can support the later stages of the verification lifecycle. In particular, it highlights opportunities to discover, align, and adapt existing formal artefacts such as contracts, specifications, and proofs. The goal is not to describe an implemented system, but to articulate an architectural vision supported by empirical insights, identifying how LLM-guided semantic inference and graph-theoretic structure could be integrated to enable scalable reuse. Our approach is that similar implementations will have similar specifications. Large repositories of verification artefacts already exist, yet their reuse remains predominantly manual. Developers must search, compare, and adapt contracts and proofs that may differ widely in syntax, abstraction level, or domain vocabulary. A long-term research challenge is therefore to design principled mechanisms that capture semantic correspondences between artefacts despite such variability. This section outlines how semantic graph representations, LLM-derived embeddings, and approximate matching algorithms could form a unified conceptual foundation for this task.

\subsection{Empirical Foundations from Knowledge-Graph and Ontology Research}

Earlier work on graph-based representations of source code explored how formal specifications could be propagated between implementations with similar structural patterns \cite{pitu2013aris}. These studies demonstrated that graph abstractions can support transfer of specifications and behavioural information across related programs. More recent developments in knowledge-graph alignment, schema matching, and entity matching provide stronger empirical support for hybrid approaches that combine symbolic structure with neural representations. Current systems show that semantic embeddings improve retrieval coverage and flexibility, while structural constraints improve precision and reduce incorrect alignments \cite{LippolisKMZJNH23}. Retrieval-augmented reasoning over graph structures has also demonstrated that grounding neural inference in symbolic representations can reduce hallucination and improve consistency.

Related evidence also comes from software engineering and ontology-engineering workflows. Code-graph retrieval systems augmented with language models demonstrate that neural methods are more effective when operating over structured representations rather than raw text alone. Similarly, ontology-guided query generation and symbolic--neural hybrid pipelines show that structural grounding and semantic interpolation are complementary capabilities \cite{vasic2025knowledge}. These observations align with our proposal that graph structure provides correctness-preserving constraints, while neural embeddings support flexible semantic interpretation across heterogeneous verification artefacts.

\subsection{Positioning Within the LIFR Framework}

Effective reuse depends on identifying subtle semantic relationships between components of verification artefacts. Programs, proofs, and specifications encode constraints among variables, transitions, invariants, and obligations. These structures rarely align through surface-level textual similarity alone, and in many cases only fragments of behavioural intent are preserved across code refactorings, platform differences, or architectural updates. For this reason, we view reuse as a problem of \emph{semantic matching under partial equivalence and abstraction}. This conceptualisation differs from contract synthesis, where the challenge is to produce new formal artefacts; here, the task is to uncover structural and semantic similarity within an existing ecosystem of artefacts. 

To support a uniform treatment of diverse artefact types, we propose modelling verification artefacts as typed, attributed graphs. Node types correspond to semantic entities--state variables, predicates, transitions, or proof obligations--while edges encode relations such as dataflow, control-flow, refinement, logical implication, or dependency. This abstraction places programs, specifications, and proofs within a shared structural representation, enabling reasoning across heterogeneous artefacts without privileging any particular syntax. The graph formalism thus acts as the symbolic substrate upon which semantic alignment can be explored.

While graph structure captures the formal relations between artefact components, it alone is insufficient for semantic matching in real-world contexts. Names, comments, logical formulas, and auxiliary documentation contain additional meaning that cannot be fully extracted via structural analysis. LLMs provide a mechanism for embedding these heterogeneous linguistic and symbolic elements into a shared semantic space. The proposal is not that LLMs determine correctness, but that they serve as semantic oracles capable of suggesting correspondences between graph nodes that are not syntactically aligned.

Within this conceptual architecture, embeddings derived from LLMs enrich graph nodes with semantic features reflecting linguistic descriptions, identifier conventions, and latent domain knowledge. Approximate matching algorithms operating on these enriched graphs could then search for candidate alignments supported both by structural consistency and by LLM-inferred similarity. This approach preserves semantic nuance without requiring the LLM to perform global reasoning or to guarantee logical soundness.

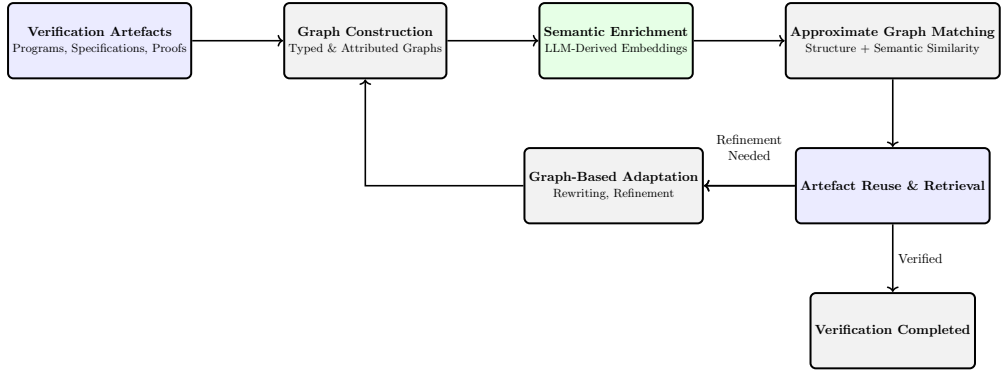
\begin{figure}[h]
\centering
\resizebox{\textwidth}{!}{%
\begin{tikzpicture}[
    node distance=1.8cm and 2.4cm,
    every node/.style={
          draw,
          line width=1.5pt,
          rounded corners,
          align=center,
          minimum height=2cm
    },
    process/.style={fill=gray!10},
    data/.style={fill=blue!8},
    llm/.style={fill=green!10},
    arrow/.style={->, very thick},
		font=\normalsize
]

\node[data] (artifacts) {\textbf{Verification Artefacts}\\
{\small Programs, Specifications, Proofs}};

\node[process, right=of artifacts] (graphs) {\textbf{Graph Construction}\\
{\small Typed \& Attributed Graphs}};

\node[llm, right=of graphs] (embeddings) {\textbf{Semantic Enrichment}\\
{\small LLM-Derived Embeddings}};

\node[process, right=of embeddings] (matching) {\textbf{Approximate Graph Matching}\\
{\small Structure + Semantic Similarity}};

\node[data, below=of matching] (reuse) {\textbf{Artefact Reuse \& Retrieval}};

\node[process, below=of reuse] (completion) {\textbf{Verification Completed}};

\node[process, left=of reuse] (adaptation) {\textbf{Graph-Based Adaptation}\\
{\small Rewriting, Refinement}};

\draw[arrow] (artifacts) -- (graphs);
\draw[arrow] (graphs) -- (embeddings);
\draw[arrow] (embeddings) -- (matching);
\draw[arrow] (matching) -- (reuse);
\draw[arrow] (reuse) -- node[right, draw=none]{\normalsize Verified} (completion);
\draw[arrow] (reuse) -- node[above, draw=none]{\normalsize Refinement\\Needed} (adaptation);
\draw[arrow] (reuse) -- (adaptation);
\draw[arrow] (adaptation.west) -| (graphs.south);

\end{tikzpicture}%
}
\caption{Visionary workflow for verification artefact reuse via hybrid graph matching and LLM-based semantic enrichment.}
\label{fig:llm-graph-reuse-workflow}
\end{figure}

Figure~\ref{fig:llm-graph-reuse-workflow} illustrates the envisioned reuse pipeline, in which existing verification artefacts are first converted into typed, attributed graphs, then semantically enriched using LLM-derived embeddings, and finally aligned through approximate graph matching that combines structural constraints with semantic similarity. LLMs would provide probabilistic semantic priors, while the graph layer would enforce global relational and type constraints. This separation of roles is essential: LLMs excel at capturing contextual semantics but are unreliable at enforcing formal structure, while graph algorithms capture structure but struggle with linguistic ambiguity.

Under this proposed architecture, reuse involves three steps:
1. \emph{Graph construction}, where artefacts are translated into typed, attributed graphs with precise relational semantics.
2. \emph{Semantic enrichment}, where each node receives an embedding, encoding textual and symbolic meaning.
3. \emph{Approximate graph matching}, where alignment candidates are discovered by combining structural similarity with embedding-level proximity.

This process differs fundamentally from contract synthesis, which concerns generating new specifications; the reuse workflow instead emphasises structural comparison, retrieval, and adaptation of existing artefacts. Identifying correspondences between verification artefacts is only one stage of reuse; adaptation is equally important. Once two specifications, proofs, or models have been aligned, portions of one artefact may need to be transformed to fit the abstraction level or structural conventions of the target context. Graph transformation theory provides a rigorous foundation for such adaptations. Existing work demonstrates how graph rewriting rules can encode insertions, deletions, refinements, and structural edits within proof assistants, preserving semantic invariants during transformation steps \cite{Strecker08GraphinProofAssitants}. We propose leveraging these ideas to express modification rules for verification artefacts. Path-expression–based views of graphs, which express relational constraints resembling those in verification logics, offer an avenue for checking whether adaptations preserve pre-established invariants. Integrated into a reuse pipeline, such transformations could enable automated restructuring of specifications or proofs while maintaining the logical relationships needed for downstream verification.

This use of LLM within an iterative reasoning cycle is becoming an accepted approach. General purpose frameworks like LangChain \cite{topsakal2023creating} and AutoGPT are gaining traction for supporting iterative development of LLM generated solutions to challenging problems. These systems support common approaches like retrieval augment generation (RAG) and especially conversational histories in a manner that's agnostic of the specific LLM being used - even allowing a consortium of multiple LLMs to be used. 

\subsection{Graph Construction and Matching for Artefact Reuse}
\label{subsec:graph-construction-matching}

Our on-going work \cite{beg2026graphstechreport} addresses the first step of artefact reuse, namely the construction of uniform representations for programs and their specifications. The study presents a pipeline that transforms imperative programs into typed, attributed graphs. The evaluation spans multiple language and specification settings, including C with ACSL, Java with JML, and Dafny for C\#. The construction process combines abstract syntax tree parsing with lightweight extraction of specification elements. These are encoded as nodes and edges capturing control flow, data dependencies, and contract-level information. The resulting graphs provide a consistent abstraction across otherwise heterogeneous artefacts. This enables comparison at a structural level while retaining key semantic elements. The work demonstrates that such graph representations can be generated systematically and at scale, forming a practical basis for further semantic analysis.

A key aspect of the approach is the integration of neural embeddings with graph structure. Node representations are enriched using models such as SentenceTransformer and CodeBERT, allowing semantic context to be captured alongside structural relations. This combination supports more expressive similarity assessment, where both program behaviour and specification intent contribute to matching. The results show that embedding-enhanced graphs improve the identification of related artefacts across datasets, even when surface syntax differs. Importantly, the approach remains modular. Graph construction and semantic enrichment are treated as separate stages, which simplifies extension and adaptation. This design also allows different embedding models or domain-specific encodings to be incorporated without changing the underlying graph structure.

Within the LIFR framework, this work provides a concrete foundation for semantic artefact reuse. It shows that reuse can be framed as a problem of alignment over structured and enriched representations rather than direct syntactic comparison. The constructed graphs act as a bridge between symbolic program structure and learned semantic similarity. This aligns with the broader goal of combining formal representations with neural methods in a controlled manner. The results also support the feasibility of approximate graph matching as a next step, where structural constraints and embedding similarity can be jointly exploited. For practitioners, the pipeline demonstrates that reusable representations can be derived from existing codebases with limited manual effort. Overall, this work strengthens the case for graph-based methods as a scalable and extensible approach to verification artefact reuse.

\section{Underlying Theory Foundations}
\label{sec:underlyingtheory}

This section develops the theoretical foundations required for semantically grounded AI assistance in specification and verification. Existing LLM-driven pipelines--such as those surveyed in \cite{BegEtAl2025, Beg2025Traceable}--typically operate at the syntactic surface of formalisms, generating constraints or predicates tied to a single verification environment such as ACSL, Event-B, or TLA$^{+}$. In the absence of a unifying semantic substrate, these artefacts lack principled interoperability, hindering cross-tool reasoning, scalable refinement, and structured reuse. Our aim is to situate LIFR workflows within mathematically robust semantic theories that enable language-independent correctness reasoning and satisfaction-preserving translations.

UTP, introduced by Hoare and He and articulated through Woodcock's tutorial \cite{Woodcock2004}, provides an algebraic relational calculus that unifies diverse programming paradigms through healthiness conditions and design theories. Its use in integrated formalisms such as Circus \cite{Woodcock2002} demonstrates its capacity to harmonise imperative constructs, CSP processes, and Z schemas within a single semantic model. Complementing UTP, the Theory of Institutions \cite{goguen1992institutions} supplies an abstract account of logical systems through signatures, sentences, models, and satisfaction, thereby enabling the comparison, translation, and combination of formalisms while preserving semantic meaning. Together, UTP and Institutions offer a rigorous, language-independent foundation for embedding AI-generated artefacts into a sound semantic framework. 

\subsection{Institutions as a Basis for Interoperability}

The theory of institutions, which finds its foundations in category theory, provides a general framework for defining a logical system \cite{goguen1992institutions}. Institutions are  used to represent logics in terms of their vocabulary (signatures), syntax (sentences) and semantics (models and satisfaction condition). The basic maxim of institutions is that ``Truth is invariant under change of notation'' \cite{goguen1992institutions}. Institutions have been devised for a number of logics \cite{sannella2012foundations} and for formal specification languages which are used for verification purposes, including CASL and Event-B \cite{mosses2004casl,farrell-rtadt-2017}. The latter is of particular interest since Event-B is used at an industrial-scale in the verification of cyber-physical systems. These institutions support modular specifications as well as interoperability between formalisms \cite{farrell2017event}. \cite{ConorPhD2023} introduces a framework in Rocq that streamlines the construction of institutions and the verification of their standard proof obligations, reducing repetitive and error-prone manual work. The approach is demonstrated through mechanised encodings of first-order logic, Event-B semantics, and their combination with linear temporal logic. The thesis also develops institution-independent constructions that support generic logic combinations and translations. 

\subsection{From Translation Frameworks to Unified Semantic Theories}

Our motivation is rooted in earlier work which relates heterogeneous formalisms through mathematically verified transformations. An example is the transformation of state-rich process algebra Circus into the state-poor process algebra CSP. The prototype developed in \cite{phdBeg16} demonstrated the feasibility of automatically translating LaTeX-based Circus specifications into CSPm, enabling direct analysis via the FDR model checker. The study revealed key technical challenges—handling complex action compositions, implementing pattern matching for behavioural normalisation, and maintaining observable event semantics—which underscored the limitations of imperative implementation techniques. These insights motivated the more principled functional formulation of translation rules later adopted. \cite{phdBeg16} also provided a comprehensive formalisation of the Circus-to-CSP translation using Haskell, encoding the refinement steps and semantic constraints as functional transformations with explicit correctness guarantees. The resulting prototype supported the formal modelling of software and hardware protocols and offered a rigorous account of how rich state-based descriptions can be systematically mapped into process-algebraic behaviours. Collectively, these contributions emphasised the need for mathematically disciplined frameworks capable of expressing and verifying relationships between disparate modelling languages—precisely the role addressed by the Unifying Theories of Programming (UTP) and the Theory of Institutions.

%

Model checking and theorem proving have long been central techniques for verifying autonomous systems. Model checkers systematically explore state spaces to identify behavioural violations, while theorem provers establish mathematical guarantees about system correctness. Existing verification frameworks combine formal behavioural models with automated reasoning to analyse timing, safety, and interaction properties in complex autonomous systems.

A significant line of work integrates process algebras with graphical state-machine notations to support robotic system design and verification. These approaches are supported by tool environments capable of automatically translating high-level models into formal representations suitable for model checking and proof-based verification. Several frameworks also combine state-machine modelling with theorem proving environments, enabling refinement-based reasoning and verification across different abstraction levels. Much of this work is grounded in the Unifying Theories of Programming (UTP), which provides a semantic basis for integrating heterogeneous formalisms through relational refinement theories. 

Mechanised UTP environments now support automated reasoning for refinement calculus, denotational semantics, and Hoare-style verification within proof assistants. Related work on model-based testing and runtime verification shows that unified semantic frameworks scale beyond theory into embedded, real-time, and industrial systems. These developments reinforce the relevance of semantic interoperability and refinement-preserving translation for future LIFR workflows. In our approach, institutional semantics extends this direction by providing a more general mechanism for interoperability across heterogeneous verification technologies and logical systems. Isabelle/UTP \cite{FOSTER2020102510} mechanises UTP theories within a proof assistant, supplying automated reasoning for refinement calculus, denotational models, and Hoare logic. Work on model-based testing \cite{6058982} demonstrates the applicability of UTP to embedded and real-time systems, while industrial deployments—from UAS traffic management with runtime verification \cite{DBLP:journals/isse/HammerCHJR22} to verification of the RTEMS operating system \cite{DBLP:conf/birthday/ButterfieldT23}—illustrate the value of unified semantic models in large-scale settings. 



Within this landscape, LIFR workflows can be re-envisioned as operating under semantic governance rather than merely generating syntactic artefacts. UTP and Institution theory allow AI-generated predicates to be checked against healthiness conditions; behavioural models to be interpreted through design theories; and specification fragments to be transported across formalisms via institution morphisms. Techniques for iterative prompting and context-sensitive refinement \cite{wang-etal-2022-iteratively} complement this vision by enabling LLMs to explore and adapt within semantically constrained spaces.

A key advantage of embedding LIFR in these semantic theories is the ability to articulate and mechanise interoperability across modelling languages and verification environments. Proof-graph frameworks such as Cyclone can encode relationships among UTP theories, Event-B machines, TLA$^{+}$ behaviours, and ACSL specifications, while satisfaction-preserving translations—e.g.\ between temporal logics or between behavioural and state-based assertions—can be expressed as institution morphisms. UTP calculi, including those mechanised in Isabelle/UTP, offer refinement, concurrency, and temporal reasoning operators compatible with automated proof engines. These mechanisms open pathways for AI-assisted migrations across formalisms, enabling structured specification reuse and multi-tool verification at scale.

Bringing these strands together, we advocate a semantically disciplined methodology for AI-assisted specification and verification in which learning systems operate within the constraints imposed by UTP and Institutional semantics. In this paradigm, AI components act not as syntactic converters but as semantic agents that propose, transform, and analyse artefacts in accordance with healthiness conditions, refinement laws, and satisfaction-preserving mappings. Such an environment enables principled cross-language verification, traceability, and scalable integration of specifications, supporting the evolution of formal artefacts beyond the limitations of tool-specific pipelines. This direction lays the groundwork for a new generation of trustworthy, interoperable, and semantically grounded verification ecosystems in which AI and formal methods co-develop in mutually reinforcing ways.

\section{Discussion}
\label{sec:discussion}

The integration of large language models into formal specification and verification pipelines exposes tightly coupled technical challenges that cannot be resolved by incremental performance improvements alone. As identified in \cite{Beg2025Traceable}, a primary limitation is the absence of aligned, high-fidelity datasets connecting real industrial natural-language requirements with semantically validated specifications and verification outcomes. Existing datasets remain synthetic, narrow, or lack verified specification-level ground truth, creating a validation gap in which LLMs may appear effective under superficial metrics while failing to preserve behavioural semantics. Future work must prioritise versioned, traceable corpora binding requirements, code, generated specifications, solver outcomes, and human corrections into a single source-aware structure. Without such datasets, true semantic correctness, refinement preservation, and scalability cannot be measured.

A second challenge is instability of specification generation under prompt variation: small phrasing differences can produce materially different ACSL or JML outputs even for fixed requirements. This undermines reproducibility in safety-critical contexts. Research must shift toward formally constrained prompt templates, contract schemas, and grammar-restricted decoding backed by tool-specific ASTs. Evaluation across prompting strategies must correlate syntactic validity with downstream proof success and solver stability. Interoperability across heterogeneous verification environments remains limited. Tool-specific translators mapping LLM outputs into ACSL, JML, or solver languages are fragile and non-scalable. A semantically typed, tool-neutral intermediate representation is needed to express preconditions, postconditions, invariants, and frame conditions independently of concrete syntax, while remaining translatable across logics with differing expressivity.

Symbolic reasoning is typically used only post hoc. Solvers detect inconsistencies after generation but do not guide it. Closed-loop integration—feeding unsatisfiable cores, counterexamples, and path constraints back into generation--requires machine-interpretable representations of solver feedback and control strategies that avoid non-terminating repair loops. Traceability across lifecycle artefacts is another bottleneck. Linking requirements, specifications, implementation, and verification results is often manual or heuristic-based and does not scale to multi-version systems. Future systems need version-aware, bidirectional trace graphs where LLMs propose candidate links validated through proof obligations or review. Semantics-preserving traceability requires refinement-aware propagation and delta-based specification updates, not full regeneration.

Handling partial, underspecified, or contradictory requirements also remains unresolved. Industrial requirements contain implicit assumptions and conflicts, and current LLMs either over-concretise or silently resolve contradictions. Generators must instead model uncertainty explicitly through conditional contracts, alternative behaviours, and assumption clauses refined during verification. Trustworthiness and certification concerns—training data provenance, memorisation, prompt injection, nondeterminism—are amplified when LLMs participate in verification pipelines. Verifiable guarantees on bounded nondeterminism, auditable training data for safety-relevant fine-tuning, and runtime monitoring of specification generation are essential for regulatory acceptance in domains such as avionics or medical software.

Scalability remains limited to small programs. Memory constraints, context-window limits, and path-coverage explosion degrade performance for larger systems. Hierarchical specification synthesis is required, where modules are verified independently and composed under formally justified and minimised interface contracts. The long-term trajectory is a shift from isolated automation to semantically grounded co-design among LLMs, symbolic engines, and human experts. LLMs should act as statistically guided front-ends under continuous logical supervision; symbolic tools must constrain generation rather than merely validate it; and human experts should curate reusable invariants, proof patterns, and failure taxonomies. The success of LLM-assisted verification ultimately depends on disciplined integration of statistical learning with formal semantic control, traceability, and certifiable trust models.

\section{Conclusions}
\label{sec:conclusions}

This paper has outlined a coherent vision for the future of formal verification centered on three tightly integrated pillars: automated contract synthesis, semantic artifact reuse via LLM-guided graph matching, and a unifying underlying theory of refinement and compositional reasoning. Contract synthesis elevates specifications from static, human-authored objects to dynamic artifacts that can be inferred, repaired, and adapted as systems evolve. At the same time, the proposed hybrid reuse framework addresses a long-standing bottleneck in formal methods: the inability to systematically transfer verification knowledge across projects and domains. By treating verification artifacts as structured semantic objects rather than isolated syntactic entities, the framework enables reuse at the level of behavioral intent rather than textual coincidence.

The integration of large language models with graph-based representations plays a critical enabling role in this vision. LLMs provide expressive semantic embeddings that bridge heterogeneous notations, languages, and abstraction levels, while graph matching enforces global structural coherence across candidate alignments. Crucially, this separation allows machine learning to guide discovery and search without compromising the symbolic guarantees required for sound verification. Rather than replacing formal reasoning, learning-based components function as semantic accelerators that expose latent structure in large verification repositories. This hybridization opens the door to verification workflows in which contracts, proofs, and counterexamples are not merely reused opportunistically, but systematically discovered, adapted, and validated at scale.

Finally, the underlying theoretical framework provides the essential glue that makes synthesis and reuse trustworthy. Refinement relations, abstraction operators, and compositional semantics supply the formal infrastructure needed to interpret approximate matches, justify reuse transformations, and ensure that transferred artifacts preserve correctness guarantees. The resulting perspective shifts formal verification from a static, one-off activity to a cumulative and continuously evolving discipline. By tightly coupling synthesis, semantic reuse, and theory, we move toward verification systems that do not merely check correctness, but actively accumulate, generalize, and propagate formal knowledge across the software and cyber-physical system lifecycle.

\section*{Acknowledgement}
This work is partly funded by the ADAPT Research Centre
for AI-Driven Digital Content Technology, which is funded by Research Ireland through
the Research Ireland Centres Programme and is co funded under the European Regional
Development Fund (ERDF) through Grant 13/RC/2106 P2. 

This publication has also emanated from research conducted with the financial 
support of Science Foundation Ireland/Research Ireland under Grant number 21/FFP-P/10118.

\section*{Use of AI-Assisted Tools}
We acknowledge the use of free version of GPT-5 for refining the textual presentation of this vision paper.
The model was applied solely to improve clarity and coherence.
The text has been thoroughly reviewed and discussed by all authors to ensure accuracy and integrity. 

\bibliography{verifai2026Bib}

\end{document}